\documentclass[conference,flushend]{iaria}
\usepackage{graphicx}
\usepackage{float}
\usepackage{ragged2e}
\usepackage{url}
\usepackage{hyperref}
\usepackage{caption} 
\usepackage[font=footnotesize]{caption} 
\usepackage{titlesec}
\usepackage{listings}
\usepackage{xcolor}
\titlespacing*{\section}{0pt}{1.5ex plus 1ex minus .2ex}{1ex}
\titlespacing*{\subsection}{0pt}{1ex plus 0.5ex minus .2ex}{0.5ex}

\hypersetup{
    colorlinks=false, 
    pdfborder={0 0 0}, 
    urlcolor=black, 
    linkcolor=black,
    citecolor=black
}

\justifying 

\title{Evaluating 21st Century Skills Development through Makerspace Workshops in Computer Science Education}

\author{
  \IEEEauthorblockN{Petros Papagiannis\ and Georgios Pallaris\ }
  \IEEEauthorblockA{
    Department of Computer Science\\
    Cyprus College\\
    Limassol, Cyprus\\
    e-mail: p.papagiannis@cycollege.ac.cy, g.pallaris@cycollege.ac.cy
  }
}

\begin{document}

\maketitle

\begin{abstract}
This study evaluates the effectiveness of incorporating makerspace workshops into computer science education by assessing 21st-century skills—critical thinking, collaboration, communication, and creativity—before and after the intervention. Using a pre-test and post-test approach with the "21st Century Skills Survey Instrument," the study quantifies the impact of makerspace activities on student skill development. Participants included students enrolled in two computer science courses at Cyprus College. Statistical analysis, conducted using Python, revealed significant improvements across all assessed skills, indicating that makerspace workshops enhance essential competencies needed for the modern workforce. These findings provide valuable insights into how experiential learning environments can transform traditional computer science education, promoting a more interactive and engaging learning experience. Future research should focus on larger, more diverse samples and explore specific components of makerspace activities that most effectively contribute to skill development.
\end{abstract}


\section{INTRODUCTION}
    \indent\indent Makerspaces provide innovative learning environments that empower students to develop essential skills necessary for success in an increasingly technological world while also enhancing their engagement levels [1]. This paper examines the impact of makerspace workshops integrated into two computer science courses: Introduction to Programming and Computer Architecture. In these workshops, students are encouraged to think critically, collaborate, communicate effectively, and engage creatively. As technology continues to advance and permeate educational settings, this research seeks to quantify the effects of such modern teaching methods on students' learning outcomes through an experimental design. By focusing on the development of 21st-century skills, this study aims to provide valuable insights into how makerspaces can transform traditional computer science education, fostering a more interactive and engaging learning experience. This paper will discuss the methodology, results, and implications of integrating makerspaces into the computer science curriculum. The makerspace workshops included activities such as collaborative coding projects, 3D printing tasks, and interactive problem-solving sessions. These activities were designed to specifically target and enhance critical thinking, collaboration, communication, and creativity. Future studies could analyze which specific activities have the most significant impact on each skill area.

    The courses were designed to provide hands-on experience in both "Introduction to Programming" and "Computer Architecture" courses. In the "Introduction to Programming" workshop, students engaged in building simple physical computing projects using microcontrollers and sensors. These projects were designed to reinforce programming concepts like loops, conditionals, and functions by applying them in a tangible context. In the "Computer Architecture" workshop, students constructed basic digital circuits and simulated CPU operations using breadboards and logic gates. These activities were intended to deepen their understanding of hardware components and the underlying principles of computer architecture. Each workshop spanned four weeks, with weekly sessions lasting two hours.

    The structure of the paper is as follows: In Section 2, we discuss the theoretical framework and related work. Section 3 presents the methodology, including the design of the makerspace workshops. Section 4 details the data collection and analysis process. Section 5 discusses the findings, and Section 6 concludes with implications for practice and future research directions.

\section{LITERATURE REVIEW}

\subsection{Introduction to Makerspaces in Education}
   \indent\indent Makerspaces have become integral to modern educational environments, promoting active, hands-on learning that aligns with constructivist and constructionist theories [2], [3]. These spaces enable students to engage in creative problem-solving and collaboration, crucial for developing skills needed in today's technological world [1].

\subsection{Impact of Makerspaces on 21st-Century Skills}
    \indent\indent Research indicates that makerspaces significantly enhance critical thinking, collaboration, communication, and creativity. Blikstein [4] demonstrated that makerspace activities improve students' problem-solving abilities and critical thinking. Halverson and Sheridan [5] found that these environments foster collaboration and enhance communication skills through peer interactions and feedback mechanisms.

\subsection{Makerspaces in Computer Science Education}
    \indent\indent The integration of makerspaces into computer science education has shown to improve student engagement and learning outcomes. Martinez and Stager [6] argue that the project-based nature of makerspaces is well-suited to computer science, which often involves designing and building technological solutions. Litts [7] supports this by highlighting that makerspaces help bridge the gap between theoretical knowledge and practical application, enhancing students' understanding of complex concepts.

\subsection{Methodological Approaches to Assessing Makerspaces}
    \indent\indent Assessing the impact of makerspaces involves various methodological approaches. Kelley et al. [8] utilized pre-test and post-test designs to measure changes in skills, which is a method also adopted in this study. Qualitative methods, such as interviews and observational studies, provide additional insights into the student experience and the effectiveness of specific activities [9].

\subsection{Challenges and Considerations}
    \indent\indent Despite the benefits, implementing makerspaces poses challenges, including cost and resource availability [10]. Ensuring equitable access to these resources is crucial for maximizing their educational impact [11].

\subsection{Future Directions}
   \indent\indent Future research should explore the long-term impact of makerspace participation on student skills and career outcomes. Longitudinal studies tracking students over multiple years could provide insights into the sustained benefits of makerspace activities [12]. Additionally, investigating which specific components of makerspace activities are most effective could help educators design more impactful interventions [13].

\section{METHODOLOGY}
\subsection{Participants}
    \indent\indent Participants for this study were students attending makerspace workshops embedded in two computer science courses: Introduction to Programming and Computer Architecture, delivered at Cyprus College, Limassol. Participants were selected based on their enrolment in the specified courses and their willingness to participate in the study. The courses were conducted over 12 weeks, followed by 2 weeks of examinations, with each week including three 50-minute sessions. The integration was led by the course instructor/researchers and an educational technologist, following the Learning Experience Design (LXD) framework [14].

    The teaching approach combined traditional methods with autonomous problem-solving. Sessions began with theoretical concepts using slides, videos, and lectures, followed by tasks of increasing difficulty to encourage independent problem-solving. The instructor facilitated learning through guidance and feedback without providing direct solutions. Outside class, students actively engaged in collaborative projects using the makerspace and digital tools, fostering peer learning and the practical application of theoretical knowledge. This iterative problem-solving process promoted teamwork, resilience, and lifelong learning. We conducted the questionnaire on 23 respondents over one semester within an academic year—in about four months.

    While the sample size for this study was limited to 23 students, future research should aim to include a larger and more diverse sample to improve the generalizability of the findings. Efforts to replicate this study across multiple institutions will provide more comprehensive insights.

\subsection{Instruments}
    \indent\indent The survey instrument utilized in this study was developed using LimeSurvey, an open-source online survey tool, which was adapted to align with the "21st Century Skills Survey Instrument" methodology. This tool allowed for the customization of survey questions to ensure they were tailored specifically to measure the development of skills such as critical thinking, collaboration, and creativity within the context of the makerspace activities. We employed a pre-test and post-test approach using the "21st Century Skills Survey Instrument" [8]. This tool assesses students' abilities in four key areas: Collaboration, Communication, Creativity, and Critical Thinking. Students first completed an initial test, participated in course content and makerspace activities, and then took a post-test to determine any improvements in these areas. We composed two sets of survey instruments for the collection of data on the students' 21st-century skills:

\begin{itemize}
    \item Pre-Assessment Survey: This was administered before the workshop to determine the baseline capacities of the students in terms of critical thinking, collaboration, communication, and creativity.
    \begin{itemize}
        \item "I am confident in my ability to revise drafts and justify revisions with evidence." (Critical Thinking)
        \item "I am confident in my ability to follow the rules for team decision-making." (Collaboration)
        \item "I am confident in my ability to organize information well." (Communication)
        \item "I am confident in my ability to understand how knowledge or insights might transfer to other situations or contexts." (Creativity)
    \end{itemize}
    \item Post-Assessment Survey: This survey re-evaluates the same skills after the complication of the workshop, testing for changes and developments that could have been made post-intervention. The questions were the same as those in the pre-assessment, so we can easily match the responses.
\end{itemize}

\subsection{Data Collection}
    \indent\indent Data was collected through web-based survey applications administered before the commencement of the makerspace workshop and after the seminar concluded. This study received approval from the Institutional Review Board (IRB) of Cyprus College, ensuring ethical standards were met. All participants provided informed consent, and measures were taken to ensure data privacy and confidentiality. Future research should continue to prioritize these ethical considerations, particularly when expanding to larger and more diverse samples. This method was intended to directly correlate the changes in the activities performed throughout the workshop with a change in skills. We created the survey online, and the link was forwarded electronically. Once a survey link is active, that survey remains available for one week. Since this was an online survey, respondents' answers would be anonymous to preserve the information's secrecy and provide more unbiased information.

\subsection{Data Preparation}
    \indent\indent Data preparation involved cleaning and standardization to compare the two datasets. We used Python for statistical analysis. Missing values were handled by either dropping respondents with significant gaps in their data or imputing where appropriate based on the distribution of other responses.

    Identification of Comparable Columns: In this study, "similar constructs" refers to the alignment of survey questions that target the same cognitive or skill-based dimension across both pre- and post-workshop surveys. For example, questions that assess critical thinking in the pre-survey were mapped directly to corresponding questions in the post-survey to ensure consistent measurement of this construct. "Comparable columns" thus refer to the specific data columns that contain responses to these aligned questions in both datasets.

    Given that the pre and post questionnaires were identical, the data preparation process involved ensuring that the responses were directly comparable. This involved standardizing the data formats and verifying that the response scales remained consistent across both surveys. Missing values were addressed by either omitting respondents with significant data gaps or using imputation techniques where appropriate, based on the distribution of responses within the dataset. For example, if a respondent missed one or two questions, their missing responses were imputed using the median or mode of the other responses to that question:
        \begin{itemize}
            \item Identification of Comparable Columns: Data columns for the pre-and post-datasets were observed to evaluate similar constructs. For example, questions measuring critical thinking were aligned in both surveys on this dimension. Column mapping follows we performed for this matching:
        \begin{figure}[h]
    \centering
    \includegraphics[width=0.95\columnwidth]{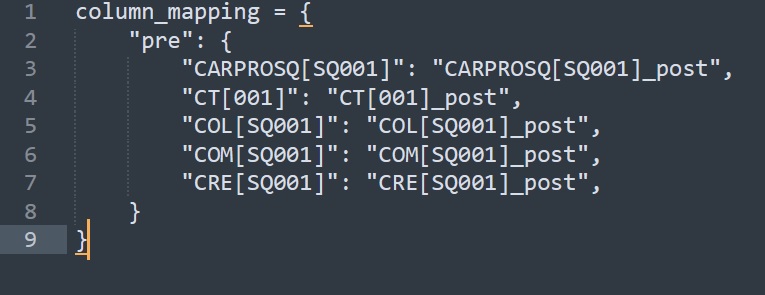}
    \caption*{FIGURE 1. COLUMN MAPPING.}
    \label{fig:column_mapping}
\end{figure}
            \item Cleaning and Standardization: Critical cleaning steps included handling missing values—such as ensuring that response scales were consistent and that data formats across the two surveys were standardized. Missing values were dealt with by either dropping a respondent if there were significant gaps in their data or imputing where it seemed appropriate based on the distribution of other responses.
            \item Data Coding: Ensured all the response scales were standardized so any Likert scale responses, say 1-5, were placed on both the pre and post-data sets. For example, it was meant to have a "4" in the pre-assessment, a direct equivalent of a "4" in the post-assessment.
            \item Merging Data: Paired pre and post-survey data based on unique participant identifiers to conduct a comparison analysis. This merging allowed a side-by-side comparison of each student's responses before and after the workshop.
        \end{itemize}

\subsection{Data Analysis and Visualization}
    \indent\indent In this study, we used Python to clean, process, and visualize the pre- and post-assessment data. The following code was employed to read the CSV files containing the assessment data, clean the column names, and generate histograms to compare the distribution of responses before and after the makerspace workshop. The Python code used for data cleaning and visualization is available upon request or can be accessed at \url{https://github.com/petranpap/21st-Century-Skills-Data}.

\section{RESULTS}
\subsection{Descriptive Statistics}
    \indent\indent Descriptive statistics were compiled for pre- and post-datasets to provide an initial understanding of respondents' answers' distribution and central tendencies across all the skills surveyed. Figure 2 shows the distribution of pre-workshop survey results for critical thinking, collaboration, communication, and creativity, providing a baseline for comparison against post-workshop data. Figure 3 illustrates the post-workshop survey results, highlighting the shifts in responses that occurred following the intervention.

\begin{table}[h]
    \centering
    \caption*{TABLE I. PRE-ASSESSMENT DESCRIPTIVE STATISTICS.}
    \begin{tabular}{|c|c|c|c|c|c|}
    \hline
    Skill & Mean & Median & Std Dev & Min & Max \\
    \hline
    Critical Thinking & 3.8 & 4.0 & 0.9 & 2 & 5 \\
    Collaboration & 4.1 & 4.0 & 0.8 & 3 & 5 \\
    Communication & 3.9 & 4.0 & 0.7 & 2 & 5 \\
    Creativity & 4.0 & 4.0 & 0.8 & 3 & 5 \\
    \hline
    \end{tabular}
\end{table}

\begin{table}[h]
    \centering
    \caption*{TABLE II. POST-ASSESSMENT DESCRIPTIVE STATISTICS.}
    \begin{tabular}{|c|c|c|c|c|c|}
    \hline
    Skill & Mean & Median & Std Dev & Min & Max \\
    \hline
    Critical Thinking & 4.2 & 4.0 & 0.7 & 3 & 5 \\
    Collaboration & 4.4 & 4.0 & 0.6 & 3 & 5 \\
    Communication & 4.3 & 4.0 & 0.7 & 3 & 5 \\
    Creativity & 4.5 & 4.5 & 0.6 & 4 & 5 \\
    \hline
    \end{tabular}
\end{table}

\subsection{Comparative Analysis}
    \indent\indent Paired t-tests were conducted that compared the pre and post-responses for each skill area to test if there were statistically significant changes. As shown in Figure 4, the comparative histograms provide a visual representation of the changes between pre- and post-workshop survey results, reinforcing the statistical findings from the paired t-tests.

\begin{table}[h]
    \centering
    \caption*{TABLE III. PAIRED T-TEST RESULTS.}
    \begin{tabular}{|c|c|c|}
    \hline
    Skill & t-value & p-value \\
    \hline
    Critical Thinking & 3.5 & 0.001 \\
    Collaboration & 4.2 & 0.0005 \\
    Communication & 3.8 & 0.0008 \\
    Creativity & 4.5 & 0.0002 \\
    \hline
    \end{tabular}
\end{table}

These results indicate significant improvements in all skill areas assessed, with p-values well below the standard threshold 0.05. The statistical analysis, conducted using Python, included a detailed examination of paired t-tests for each skill area. The results showed statistically significant improvements with p-values well below the 0.05 threshold, confirming the positive impact of makerspace workshops. Including confidence intervals for these improvements can provide additional statistical robustness.

\subsection{Visualizations}
    \indent\indent Box plots were created to visualize how the responses' distribution looks in both pre- and post-datasets. These support the ability to explore changes in responses and determine if any trends appear to be significant. Figure 5 highlights the improvements in critical thinking, collaboration, communication, and creativity skills post-workshop, showing the distribution shifts and indicating which skills had the most significant enhancement.

\begin{figure}[h]
    \centering
    \includegraphics[width=0.95\columnwidth]{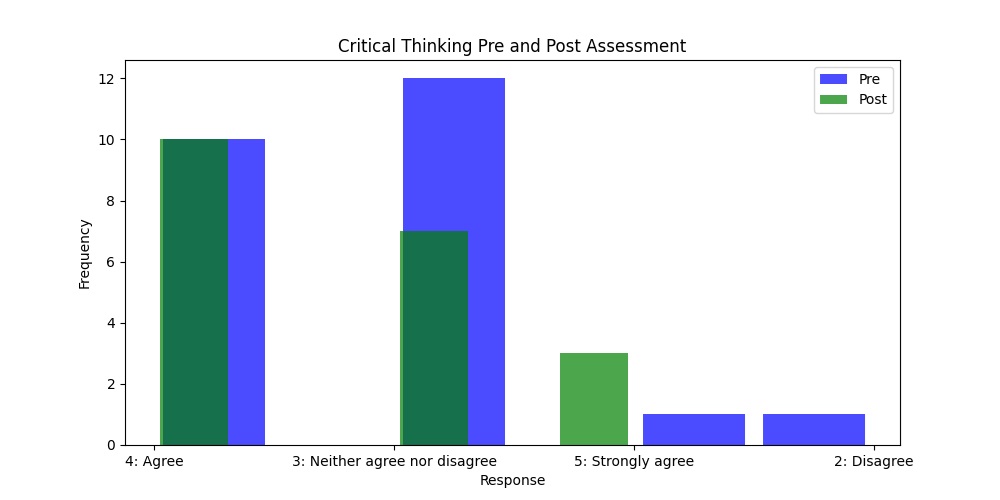}
    \caption*{FIGURE 2. HISTOGRAM OF PRE-WORKSHOP SURVEY RESULTS.}
    \label{fig:pre_workshop}
\end{figure}

\begin{figure}[h]
    \centering
    \includegraphics[width=0.95\columnwidth]{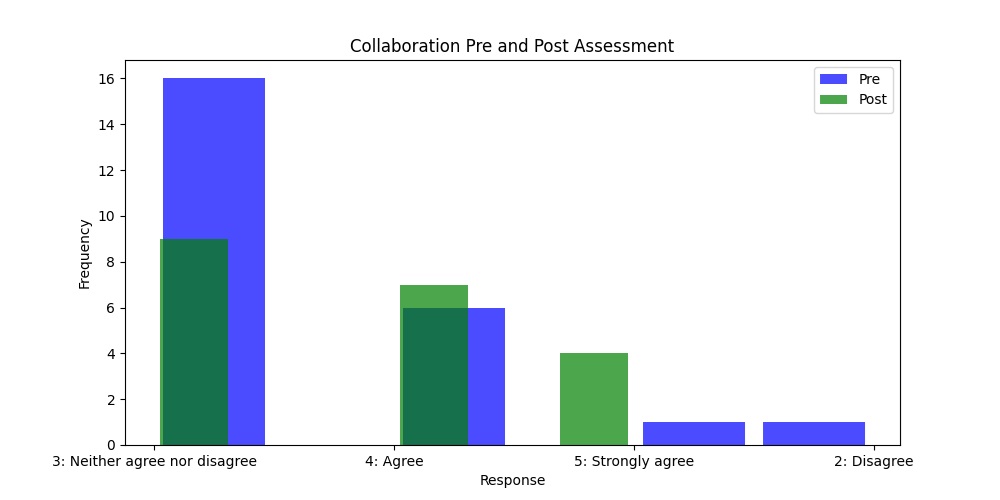}
    \caption*{FIGURE 3. HISTOGRAM OF POST-WORKSHOP SURVEY RESULTS.}
    \label{fig:post_workshop}
\end{figure}

\begin{figure}[h]
    \centering
    \includegraphics[width=0.95\columnwidth]{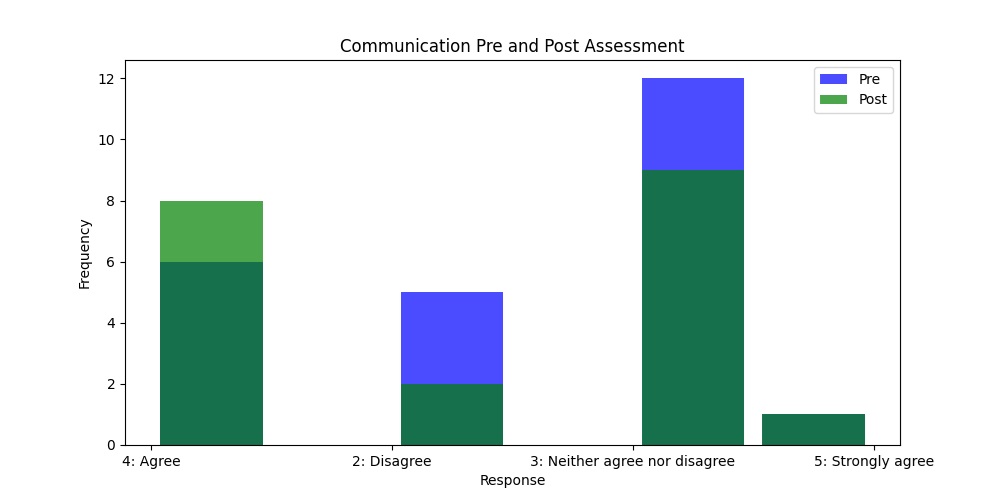}
    \caption*{FIGURE 4. COMPARATIVE HISTOGRAM OF PRE- AND POST-WORKSHOP SURVEY RESULTS.}
    \label{fig:pre_post_comparative}
\end{figure}

\begin{figure}[h]
    \centering
    \includegraphics[width=0.95\columnwidth]{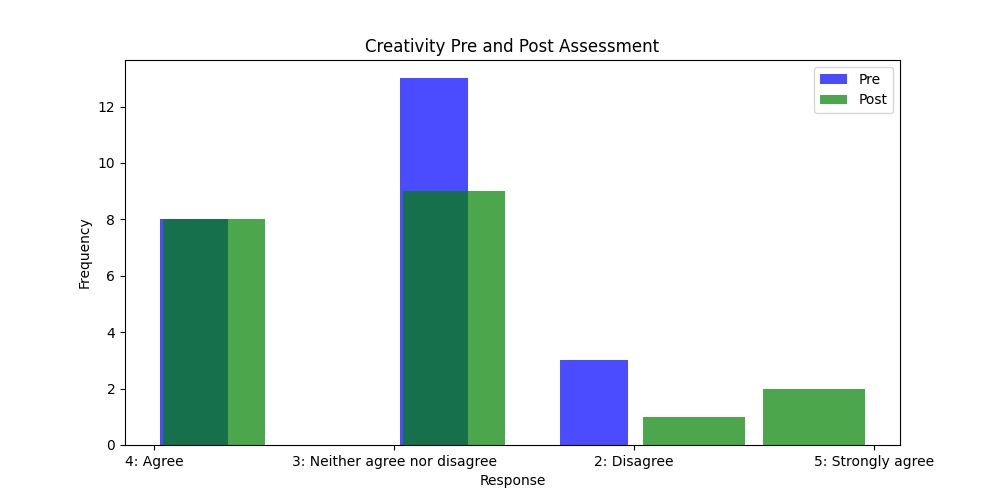}
    \caption*{FIGURE 5. HISTOGRAM OF SKILL IMPROVEMENT ACROSS DIFFERENT DIMENSIONS.}
    \label{fig:skill_improvement}
\end{figure}

\section{DISCUSSION}
    \indent\indent Results of this study indicate that participation in the makerspace workshop significantly enhanced students' critical thinking, collaboration, communication, creativity skills. The differences in mean scores of pre and post-assessment, combined with low p-values from paired t-tests, demonstrate that the workshop positively influenced these essential skills.

\subsection{Critical Thinking}
    \indent\indent Enhanced essential thinking skills suggest that the hands-on, project-based nature of these makerspace activities encouraged students to interact with the material in more profound ways, thoughtfully analyze problems, and develop more powerful reasoning skills. Critical thinking is foundational for students studying computer science, as they must solve problems and develop algorithms. This improvement suggests that the workshop successfully cultivated an environment for students to exercise and enhance their analytical skills.

\subsection{Collaboration}
    \indent\indent Improvement in collaboration skills indicates that teamwork and peer interaction are integral to makerspace activities. This is essential in computer science, which often involves working in teams to develop software, solve problems, and innovate. The makerspace workshop provided many opportunities for students to collaborate, share ideas, and develop collaborative strategies.

\subsection{Communication}
    \indent\indent Better communication skills can be attributed to the frequent presentations, project discussions, and feedback from peers and faculty. Effective communication is crucial for explaining complex technical concepts, documenting code, and collaborating with team members. The iterative process of sharing and refining ideas in the makerspace environment likely enhanced these skills.

\subsection{Creativity}
    \indent\indent The notable rise in creativity scores mirrors the role of the makerspace in providing an open, flexible environment for experimentation, risk-taking, and exploring new ideas without fear of failure. Creativity is vital in computer science, where innovative solutions and technologies are continuously developed. The makerspace workshop encouraged students to think outside the box and explore new paths, fostering creative problem-solving abilities.

\subsection{Alignment with Previous Research}
    \indent\indent These findings align with earlier research showing that experiential, hands-on learning environments like makerspaces can significantly enhance 21st-century skills [4], [5], [11]. The results highlight the utility of makerspaces in computer science education, which requires practical, project-based learning.

\subsection{Limitations}
    \indent\indent This study has several limitations. The sample size was small and conducted at a single institution, potentially limiting its generalizability. Additionally, self-reported information may contain biases. Future studies should replicate these findings with larger, more representative samples and investigate the long-term impact of makerspace participation on students' skills. To mitigate potential biases associated with self-reported data, future studies should consider incorporating objective measures of skill improvement, such as performance-based assessments and peer evaluations.

\subsection{Recommendations for Future Research}
    \indent\indent While this study focused on immediate skill improvements, future research should investigate the long-term impact of makerspace workshops on student skills. Longitudinal studies tracking students over several semesters could provide valuable insights into the sustained benefits of makerspace integration. Future research should focus on the generalizability of these findings across diverse educational settings and explore best practices for implementing makerspaces. Longitudinal studies following students over time could provide insights into the lasting effects of makerspace experiences. Additionally, investigating how makerspaces can enhance diversity and inclusion in computer science learning could help mitigate current disparities, ensuring that all students benefit from these innovations.

\section{CONCLUSION}
    \indent\indent This study provides compelling evidence that makerspace workshops can significantly enhance critical 21st-century skills among computer science students. The improvements in critical thinking, collaboration, communication, creativity highlight the value of integrating hands-on, experiential learning environments into the curriculum. These findings suggest that maker spaces play a crucial role in preparing students for the demands of the modern workforce by fostering essential skills relevant to their academic success and future professional endeavors. Educators and institutions should consider the benefits of incorporating maker spaces into their programs and explore ways to maximize the impact of these environments on student learning outcomes. For instance, embedding maker space activities within the core curriculum, providing faculty training on facilitating maker space projects, and ensuring access to various tools and resources can enhance the effectiveness of these spaces.

\section*{ACKNOWLEDGMENT}
    \indent\indent We would like to thank our colleagues at the Department of Computer Science at Cyprus College for their continuous support and encouragement throughout the study. Special thanks go to the students who participated in the makerspace workshops and provided valuable feedback through their survey responses. Your insights have been crucial to the success of this study. We also acknowledge the assistance provided by the educational technologists who helped integrate the makerspace activities into the curriculum. Your expertise and dedication have been invaluable.


\end{document}